# That's C, baby. C!

## And there is nothing you can do with it without the build. Nothing!

*Roberto Bagnara,* BUGSENG[*] and University of Parma[†] member[‡] of MISRA C Working Group, member[‡] of ISO/IEC JTC1/SC22/WG14 - C Standardization Working Group

Hardly a week goes by at BUGSENG without having to explain to someone that almost any piece of C text, considered in isolation, means absolutely nothing. The belief that C text has meaning in itself is so common, also among seasoned C practitioners, that I thought writing a short paper on the subject was a good time investment. The problem is due to the fact that the semantics of the C programming language is not fully defined: non-definite behavior, predefined macros, different library implementations, peculiarities of the translation process, ...: all these contribute to the fact that no meaning can be assigned to source code unless full details about the build are available. The paper starts with an exercise that admits a solution. The existence of this solution will hopefully convince anyone that, in general, unless the toolchain and the build procedure are fully known, no meaning can be assigned to any nontrivial piece of C code.

## Exercise

Write a C function that returns an integer to the caller, with the following constraints:

1. the function source code text is self-contained (no access to external code, including the standard library) and is shorter than 700 characters;

2. the function has no state and no input whatsoever (no static variables, no parameters, no access to globals, no access to the environment);

3. the function uses only the features of the C language specified in any version of the ISO C standard, its output does not depend on unspecified or undefined behavior, and its code does not exceed any minimum implementation limit;

4. the function can be compiled in more than 700 different ways, each corresponding to a single invocation of GCC/x86_64 (version 8 or later), so that it returns more than 700 different values;

5. the used GCC command lines do not include any `-D` or `-U` options (including synonyms, i.e., no explicit fiddling with macros), use no assembler options (i.e., no fiddling with assembly code), no linker options (i.e., no fiddling with object files and libraries), and no directory options (i.e., tools, header files and libraries will only be searched in standard places).

A solution will be given later in the paper. But

> "do not turn to the answer until you have made a genuine effort to solve the problem by yourself, or unless you absolutely do not have time to work this particular problem. After getting your own solution or giving the problem a decent try, you may find the answer instructive and helpful."
>
> (Donald E. Knuth, TAOCP)

---

[*] mailto:roberto.bagnara@bugseng.com
[†] mailto:bagnara@cs.unipr.it
[‡] Writing and speaking in a personal capacity.

# C Is Not Fully Defined

C is defined by international standards: it was first standardized in 1989 by the American National Standards Institute (this version of the language is known as ANSI C) and then by the International Organization for Standardization (ISO) [5–9].

For very good reasons [1], the C language is not fully defined. The concrete possibility of generating very efficient code even with relatively simple compilers on any architecture is the main factor why we have implementation-defined behavior: *unspecified behavior where each implementation documents how the choice is made* [9, Par. 3.4.1]; e.g., the sizes and precise representations of the standard integer types;

undefined behavior: *behavior, upon use of a nonportable or erroneous program construct or of erroneous data, for which this International Standard imposes no requirements* [9, Par. 3.4.3]; e.g., attempting to write to a string literal constant;

unspecified behavior: *use of an unspecified value, or other behavior where this International Standard provides two or more possibilities and imposes no further requirements on which is chosen in any instance* [9, Par. 3.4.4]; e.g., the order in which function call arguments are evaluated.

The last two kinds of *non-definite behavior* (this is how we collectively call all instances of behavior that is not fully defined), when present, make the semantics of a program completely unpredictable. We will not concern ourselves with these kinds of behavior in this paper.

Another kind of behavior was added to the language in order to make it universally useful:

locale-specific behavior: *behavior that depends on local conventions of nationality, culture, and language that each implementation documents* [9, Par. 3.4.2]; e.g., character sets and how characters are displayed.

Locale-specific behavior influences, for instance, the reading and writing of integer and floating-point numbers, the formatting of dates, string sorting and so on. We will disregard this kind of behavior as well.

## Implementation-Defined Behavior

Due to implementation-defined behavior, calling C *a programming language* is not quite accurate. In reality C is *a huge family* of languages, each of which can be called a *dialect of C*; in the C standard, a dialect of C is referred to as a *conforming implementation*.

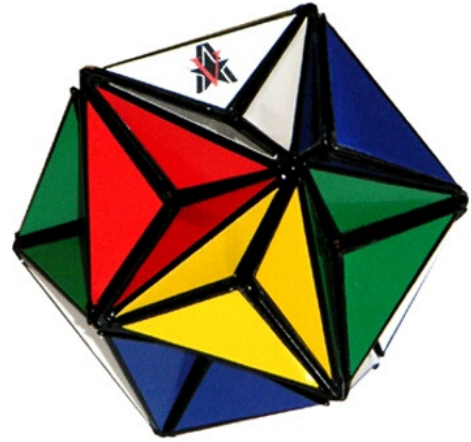

Figure 1: *Alexander's Star: $7.24 \times 10^{34}$ positions*

Let us consider C99 [7], which is still the most used version of C for the development of embedded systems. In C99, there are 112 implementation-defined behaviors. By definition, each implementation-defined behavior can be defined in 2 or more ways. This implies that the number of languages that can be legitimately called "C" is greater than $2^{112} \approx 5.19 \times 10^{33}$. Actually, choosing integer types in $\{8, 16, 32, 64\}$ respecting the constraints of C99 brings us to more than $10^{35}$ possible dialects. For comparison, Alexander's Star, a puzzle similar to the Rubik's Cube, in the shape of a great dodecahedron, has a smaller number of distinct positions: see Figure 1.

Generally speaking, a given compiler can implement, via options, several such dialects of C. For an instance, GCC/x86_64 —the compiler mentioned in the exercise opening this paper—, implements, via options, *thousands of dialects of C*. On the other hand, GCC/x86_64, does not implement any dialect in which a `char` is 16 bits: Texas Instruments' TMS320C28x optimizing C/C++ compiler does [13]; and the Freescale/CodeWarrior compiler for HC(S)12 allows selecting, via command-line options, the sizes of all integer and floating-point types [3]. Only by taking into account the compiler options that are actually used in the build, can we assign a meaning to a C program. This is further complicated by the fact that compiler options may be given:

- on the command line;
- in environment variables;
- in configuration files.

Note also that different translation units can be compiled with different options.

## There Is No Easy Solution

While it is possible to write C code that reduces the dependencies on implementation-defined behavior, this is harder than most people would expect. MISRA C:2012 [11] goes a long way in this direction, and it also prescribes documentation of any implementation-defined behaviors that are relevant to the program at hand, but this is not sufficient. For instance, many people believe that, just by using the types of `<stdint.h>` instead of, say, `short` and `int`, the program will be immune from the implementation-defined behavior depending on the sizes of basic types. Unfortunately, this is not true. Consider:

```
#include <stdbool.h>
#include <stdint.h>

bool add_overflow(uint16_t x,
                  uint16_t y) {
  return (x + y) < x;
}
```

Overflow will *not* be detected on machines where the size of `int` is larger than 16 bits.[1] To make the function work as expected on all machines, it should be written as follows:

```
#include <stdbool.h>
#include <stdint.h>

bool add_overflow(uint16_t x,
                  uint16_t y) {
  return (uint16_t)(x + y) < x;
}
```

There are other aspects that compound to the number of C dialects: predefined macros and algorithms for searching include files.

## Predefined Macros

Every implementation can (and usually does) provide predefined macros: such macros may begin with an underscore, and thus be part of the implementation name space, or not begin with an underscore, and thus be part of the user name space [9, Par. 6.10.8]. Often the predefined macros are functions of the compiler options. For instance, Microchip's MPLAB C18 C Compiler predefines macro `__18F258` to the constant `1` if the command-line option `-p18f258` was given [10].

---
[1] Note also that the returned expression does not violate any MISRA C guideline.

## Included Files

Another implementation-defined aspect of C that adds considerable complication concerns the selection of header files that are the subject of `#include` directives [9, Par. 6.10.2]. Places that may or may not be searched by compilers, in orders that differ from compiler to compiler and that may be influenced by compiler options, involve:

- whether the header is specified with `< >` or `" "`;
- paths relative to the compiler executable;
- the path of the main file;
- the paths of the direct and/or indirect includers;
- the current working directory of the compiler process;
- paths specified (directly or indirectly) by compiler options.

Taking into account predefined macros and algorithms for searching included files, it is clear that the number of C dialects is much more than our previous rough estimation suggested.

## Consequences for Static Analysis

The huge variety in C dialects implies that static analysis tools that are supposed to work with multiple C dialects must:

1. Specify which sets of dialects are supported, at least in terms of the supported toolchains.

2. Provide practical ways to obtain full information about the build to be analyzed: not just the source files, but also which compiler (and linker, and librarian) are used and with which options.

3. Adapt the analysis to the build.

In other words, someone must adapt the tool execution to the particular dialect implemented by *that compiler* with *that set of options*, possibly *for each translation unit* that composes the system being analyzed. Who is that "someone"? It can be:

- The user, who must be aware of the fact that properly taking into account all relevant peculiarities of the build is a daunting and very error-prone task. And, of course, it requires a tool that supports these variations. Moreover, if anything changes in the build procedure (e.g., one compilation option), tool configuration must be adapted correspondingly.

- The tool itself: this is much, much better for the user. But, of course, considerable sophistication is required in the tool design and implementation. Moreover, the tool must be kept up to date with the evolution of the toolchains it supports.

The importance of all this cannot be stressed enough: we have seen many projects whose MISRA C compliance [11] was completely led astray due to misconfiguration of the actual language dialect implemented by the build being analyzed, and this in safety-critical projects. One might argue that for some simple rules, analysis can be reliably conducted without taking into account the language dialect. This is not so: if you consider predefined macros and the algorithm used to search header files, you realize that not capturing those correctly means that, in general, you would be analyzing code that is different from the code that has been compiled and embedded into the device.

Here is an example of how this fatal mistake might happen: we have code implementing two versions of a the (supposedly) same dispatch table; a fast one based on GCC's computed gotos[2] and a standard one based on a `switch` statement. The intention is to use the former only with GCC and when optimization is enabled and to use the latter in all other cases; the two cases being discriminated by the predefined macros `__GNUC__` and `__OPTIMIZE__`. It is clear that, if predefined macros are not correctly captured, we cannot reliably check compliance with respect to the MISRA C guidelines restricting the use of `goto` statements and the use of compiler extensions.

## Answer to the Exercise

The following C function source code is 647 characters long. The function can return all integer values from 0 to 767, depending on how it is compiled with GCC/x86_64 (version 8 or later).

```c
unsigned return_value(void) {
  typedef enum { Z } S;
  typedef enum {
    I = ((int)-1U/2 == (int)(-1U/2))
        ? (int)-1U : (int)(-1U/2)
  } L;
  struct {
    int f:8;
  } s = { 255 };
  unsigned m = 0;
  m += (((char)-1) < 0) ? 1 : 0;
  m += (s.f < 0) ? 2 : 0;
  m += (sizeof(S) < sizeof(L)) ? 4 : 0;
#ifdef __OPTIMIZE__
  m += 8;
#endif
  m += (sizeof(void *) == 8)
       ? 16 : 0;
#if !defined(__STDC_HOSTED__) \
    || __STDC_HOSTED__ == 0
  m += 32;
#endif
#ifdef __STDC_VERSION__
  m += (__STDC_VERSION__ % 4)*64;
#else
  m += 192;
#endif
#ifndef __STRICT_ANSI__
  m += 256;
  m += (sizeof("??-") != 4) ? 256 : 0;
#endif
  return m;
}
```

For example, you can obtain

- 0, with `-funsigned-char -funsigned-bitfields -m32 -fhosted -std=c11`;
- 1, with `-fsigned-char -funsigned-bitfields -m32 -fhosted -std=c11`;
- 7, with `-fsigned-char -fsigned-bitfields -fshort-enums -m32 -fhosted -std=c11`;
- 42, with `-funsigned-char -fsigned-bitfields -O2 -m32 -ffreestanding -std=c11`;
- 443, with `-fsigned-char -fsigned-bitfields -O2 -m64 -ffreestanding -std=gnu17`;
- 640, with `-funsigned-char -funsigned-bitfields -m32 -fhosted -std=gnu17 -trigraphs`.

Source code and supporting scripts that allow reproducing this experiment, both under Windows and Linux are available at https://bugseng.com/that-is-c-baby-c. There is a main program that can be invoked in two ways:

1. with a numeric argument between 0 and 767, the program prints the options to be given to GCC in order to obtain the specified value;

2. without any argument, the program prints the value returned by `return_value()`.

---
[2] A GCC extension whereby one can get the address of a label defined in the current function, store it, and then use it as the target of `goto` statements.

In addition, there are scripts `test.bat` and `test.sh` to automatically check all the possible values. Here is a sample session, where backslashes mark newlines that are in the paper and not in the actual input/output and option `-w` is used to suppress a warning concerning trigraphs:

```
$ gcc -w main.c
$ a.exe 100
-funsigned-char -funsigned-bitfields \
 -fshort-enums -m32 -ffreestanding -std=c99
$ gcc -funsigned-char \
 -funsigned-bitfields -fshort-enums -m32 \
 -ffreestanding -std=c99 main.c
$ a.exe
100
$ test.bat
tests 0..767 succeeded
```

# Conclusion

Theoretically speaking, C can be seen as an infinite family of C dialects: that big is the latitude the C standards allow to conforming implementations. In practice, considering the number of compilers in use today and the combinations of implementation-defined behaviors they support, the number of C dialects in use is a finite but large number (on the order of tens of thousands).

In this paper, we have shown how failure to capture the C dialect for which a piece of C code is intended makes it impossible, in general, to draw any conclusion about its behavior. As a consequence, no analysis tool can draw any reliable conclusion about a piece of C code unless all relevant information is available. Unfortunately, in industry, including in safety-critical sectors, there is little awareness of this fundamental property of C: many practitioners seem to believe that they can use a static analysis tool without worrying about the fact that their compilation toolchain and the analysis tool agree on the implementation-defined aspects of the language. In this crucial mistake, they are not helped by tool vendors bragging about so-called "tool certificates": upon close inspection it can often be seen that neither the tool nor the "certificate" pay sufficient attention to this matter. It is not by chance that functional safety standards, such as CENELEC EN 50128 [2], ISO 26262 [4] and RTCA DO-178C [12], clearly state that, while applications can be *certified* for use, tools employed in their development or verification can only be *qualified* in the specific context of their use: for a C static analysis tool, such context includes the C dialects that are actually used in the different parts of the project.

# Acknowledgments

I am grateful to the following people who provided useful comments and advice on previous versions of this paper: Abramo Bagnara (BUGSENG), Paolo Bandini (University of Parma and BUGSENG), and Patricia M. Hill (BUGSENG), Marcel Beemster (Solid Sands).